\documentclass[sigconf]{aamas} 

\usepackage{balance} 
\usepackage{algorithm}
\usepackage[noend]{algpseudocode}
\usepackage{amsfonts, amsthm}
\usepackage{mathtools}
\usepackage{bm}
\usepackage{cleveref}
\usepackage{subcaption}
\usepackage{multirow}

\doi{TNEX7143}

\makeatletter
\gdef\@copyrightpermission{
  \begin{minipage}{0.2\columnwidth}
   \href{https://creativecommons.org/licenses/by/4.0/}{\includegraphics[width=0.90\textwidth]{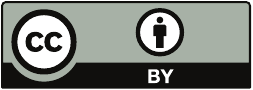}}
  \end{minipage}\hfill
  \begin{minipage}{0.8\columnwidth}
   \href{https://creativecommons.org/licenses/by/4.0/}{This work is licensed under a Creative Commons Attribution International 4.0 License.}
  \end{minipage}
  \vspace{5pt}
}
\makeatother

\setcopyright{ifaamas}
\acmConference[AAMAS '26]{Proc.\@ of the 25th International Conference
on Autonomous Agents and Multiagent Systems (AAMAS 2026)}{May 25 -- 29, 2026}
{Paphos, Cyprus}{C.~Amato, L.~Dennis, V.~Mascardi, J.~Thangarajah (eds.)}
\copyrightyear{2026}
\acmYear{2026}
\acmDOI{}
\acmPrice{}
\acmISBN{}

\acmSubmissionID{933}

\title{Zero-Shot Coordination in Ad Hoc Teams with Generalized Policy Improvement and Difference Rewards}

\author{Rupal Nigam}
\affiliation{
  \institution{University of Illinois Urbana-Champaign}
  \city{Champaign, IL}
  \country{United States}}
\email{rupaln2@illinois.edu}

\author{Niket Parikh}
\affiliation{
  \institution{University of Illinois Urbana-Champaign}
  \city{Champaign, IL}
  \country{United States}}
\email{niketnp2@illinois.edu}

\author{Hamid Osooli}
\affiliation{
  \institution{University of Illinois Urbana-Champaign}
  \city{Champaign, IL}
  \country{United States}}
\email{hosooli2@illinois.edu}

\author{Mikihisa Yuasa}
\affiliation{
  \institution{University of Illinois Urbana-Champaign}
  \city{Champaign, IL}
  \country{United States}}
\email{myuasa2@illinois.edu}

\author{Jacob Heglund}
\affiliation{
  \institution{University of Illinois Urbana-Champaign}
  \city{Champaign, IL}
  \country{United States}}
\email{jheglun2@illinois.edu}

\author{Huy T. Tran}
\affiliation{
  \institution{University of Illinois Urbana-Champaign}
  \city{Champaign, IL}
  \country{United States}}
\email{huytran1@illinois.edu}

\begin{abstract}
Real-world multi-agent systems may require ad hoc teaming, where an agent must coordinate with previously unseen teammates in a zero-shot manner. Prior work either selects a pretrained policy based on an inferred model of the new teammates or pretrains a single policy that is robust to potential teammates. Instead, we propose to \textit{dynamically leverage all pretrained policies} through two key ideas---generalized policy improvement and difference rewards---for efficient and effective knowledge transfer between different teams. We empirically demonstrate that our algorithm, Generalized Policy improvement for Ad hoc Teaming (GPAT), successfully enables zero-shot transfer to new teams in three simulated environments and demonstrate our algorithm in a real-world multi-robot setting.
\end{abstract}


\newcommand{\BibTeX}{\rm B\kern-.05em{\sc i\kern-.025em b}\kern-.08em\TeX}
\newcommand{\learner}{\text{a}}
\newcommand{\defeqn}{\coloneqq}
\DeclareMathOperator*{\argmax}{argmax}

\theoremstyle{definition}
\newtheorem{definition}{Definition}[]
\theoremstyle{problem}
\newtheorem{problem}{Problem}[]

\begin{document}


\pagestyle{fancy}
\fancyhead{}


\maketitle 


\section{Introduction}
\label{sec:intro}

Ad hoc teaming (AHT), where an autonomous agent must coordinate with other unknown agents \cite{stone_ad_2010}, is an open challenge for multi-agent systems. Consider a search-and-rescue mission where robots are deployed from different organizations and expected to coordinate with each other on the fly---these robots may have different biases in how they achieve an objective (e.g., risky vs. risk-averse) or have different capabilities (e.g., sensing vs. manipulation). Adapting to such differences would enable agents to effectively and autonomously complete tasks where the team is unknown prior to deployment. We therefore focus on zero-shot coordination (ZSC) for AHT, where the controlled agent, or the learner, is able to pretrain with various teams but then must coordinate with a new team with no online learning \cite{mirsky_survey_2022}.

Type-based approaches are common for AHT, where the learner pretrains with a set of teammate types, then infers the best pretrained policy to use with the new team at test time  \cite{BARRETT-plastic,li_individualized_2021,zhao_coordination_2022}. However, these methods struggle to handle new teams not seen during pretraining and require online inference of the new team type. An alternative approach pretrains a learner that is robust to new teams through careful generation of diverse pretraining teams \cite{hu_other-play_2020, zhao_maximum_2022, yu2023learning}. However, these methods often require large training populations, may suffer from overfitting, and can struggle to generalize to out-of-distribution teammates \cite{wang2024zsc}. Generating many training teammates may also be infeasible for real-world applications, for example, due to computational constraints.

We address these challenges through two key ideas. Our first idea is to leverage a library of pretrained learner policies, but instead of choosing one at test time based on the inferred team type, we dynamically leverage the whole library with \emph{no online inference or learning.} We specifically use a \emph{generalized policy improvement (GPI)} policy to select from pretrained policies given the current state, motivated by the fact that GPI can guarantee improvement over library policies \cite{barreto_transfer_2018}. However, this guarantee is only valid when the dynamics are constant and the reward function differs---for AHT, dynamics now change due to new teammate behaviors, while the reward stays the same.
Our second idea is then to use \emph{difference rewards} to define the value functions of pretrained policies used by a GPI policy. Difference rewards address the multi-agent credit assignment problem by approximating the contribution of an individual agent to a team reward \cite{proper2012}. We use this idea to reduce the impact of the distribution shift induced by new teammates. 

We integrate these ideas into an algorithm for ZSC in AHT and empirically demonstrate the benefits of our method.
We summarize our contributions as follows:
\begin{enumerate}
    \item we propose an algorithm for ZSC in AHT based on GPI with difference rewards (\Cref{sec:method}),
    \item we empirically demonstrate the benefits of our method relative to baselines in three simulated environments and demonstrate its use in a multi-robot system (\Cref{sec:experiments}).
\end{enumerate}


\section{Related Work}
\label{sec:sota}
Type-based approaches to AHT use a pretrained library of policies, where the learner selects an appropriate response based on inferred teammate behavior. For instance, \cite{BARRETT-plastic} updates a prior over the pretrained policies through online inference with the new teammate to determine the most likely teammate type, and acts using the corresponding learner policy. 
Similarly, \cite{li_individualized_2021} and \cite{ni_adaptive_2021} infer a teammate’s policy using a similarity metric and select a complementary learner policy to coordinate in Team Space Fortress.
\cite{zandOntheFly2022} leverages Gibbs sampling to update a distribution over possible learner policies in Hanabi and similarly use this distribution to select a pretrained policy.
\cite{zhao_coordination_2022} extends these methods by employing a mixture-of-experts approach to mix pretrained policies in the Overcooked environment, rather than selecting a single policy. Their method first identifies behaviors through unsupervised clustering of previously collected data, trains learner policies for each behavior, and then uses online samples to update belief weights over each policy for a mixture-of-experts model.
However, these methods require observations with the unseen teammate to appropriately infer the best response from the pretrained library, and thus are not truly zero-shot.
Many of them also implement a single pretrained policy, limiting their ability to combine pretrained skills at execution.

Another class of methods focuses on training a single robust policy by generating a diverse training pool of teammates for ZSC.
\cite{hu_other-play_2020} achieves this by breaking symmetries inherent in the task through random relabeling of the teammate's states and actions. \cite{strouse2021collaborating} generates a diverse set of training teammates by using different random seeds and checkpoints during the training of agents. \cite{yu2023learning} expands this concept by searching over the reward space to construct a training pool. \cite{zhao_maximum_2022} enhances diversity in training teammates using an entropy bonus and employing a prioritization strategy to select training teammates. Lastly, \cite{lucas2022any} ensures diversity in teammate policies by considering intrinsic rewards with random navigation.
However, these methods can also be prone to overfitting and may struggle to generalize to teammates outside of the training pool \cite{wang2024zsc}.
Additionally, many of these methods require large training pools, resulting in high computational costs.
Finally, it may not be possible to define pretraining teammates in certain real-world settings, due to, for example, limited resources and access to robot platforms. Instead, it may be more practical in certain settings for a few representative teammates to be given to the learner.

\begin{figure}[t]
    \centering
    \includegraphics[width=\linewidth]{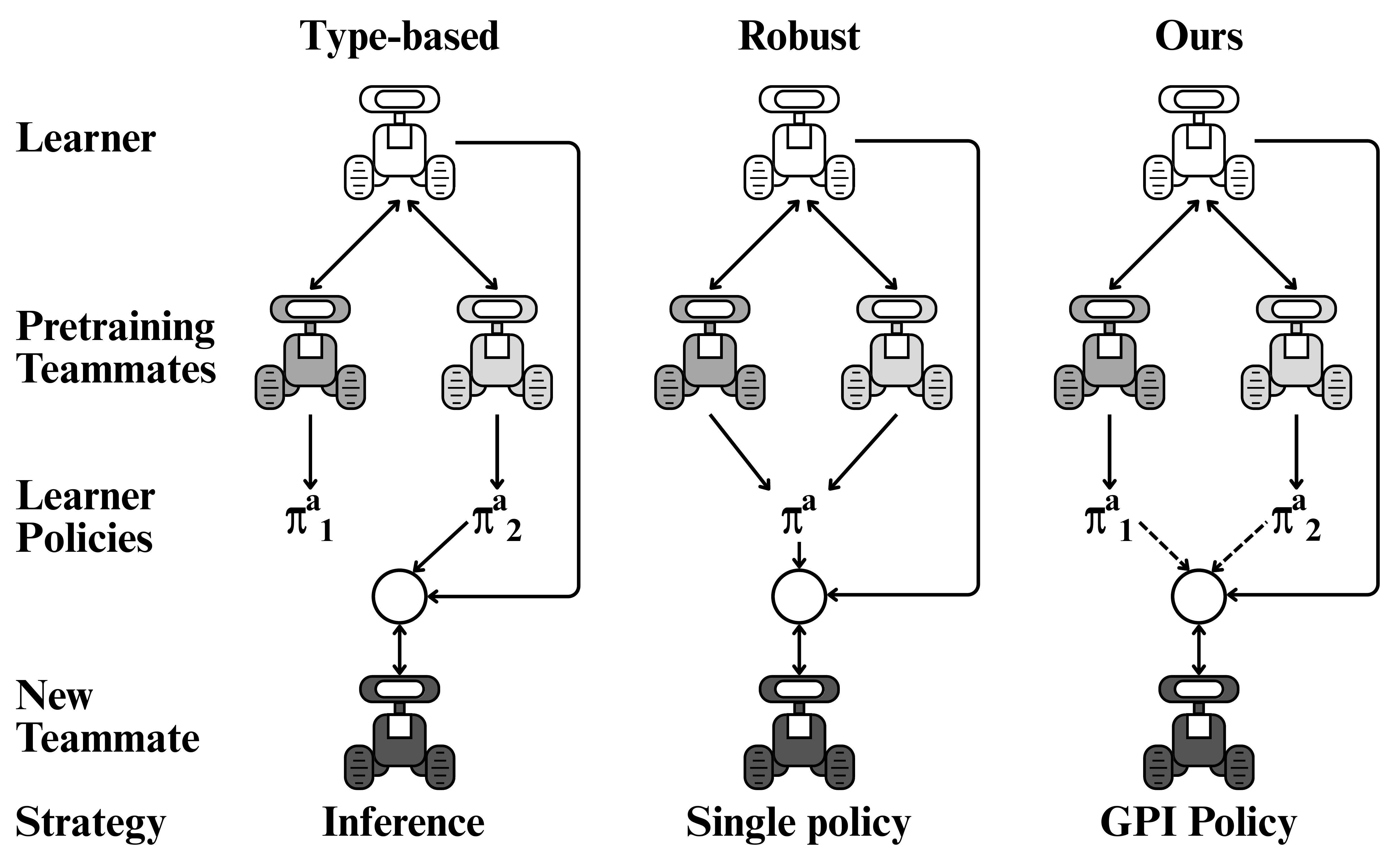}
    \caption{Common methods for AHT either select a pretrained policy based on the inferred behavior of the teammate (Type-based) or train a single policy that is robust to potential teammates by generating a diverse training pool (Robust). We instead dynamically select which policy to use at every timestep without inference for ZSC.}
    \label{fig:aht-methods}
\end{figure}

We illustrate these two common approaches for AHT in \Cref{fig:aht-methods}, alongside our proposed approach. Other AHT approaches are complementary to this paper. \cite{gu2022online} and \cite{ribeiro2022assisting} address AHT with the additional challenge of partial observability, where teammate actions and environment rewards are unobservable, so the learner updates its belief prior over the library using online observations.
\cite{mirsky2020} investigates the benefit of communication between teammates, while in this work we assume no communication between agents.
\cite{nekoeiFewshotCoordinationRevisiting2023} investigates the adaptation ability of AHT approaches in a few-shot, rather than zero-shot, coordination setting.
\cite{rahman2021} leverages graph structures to handle ad hoc teams where agents can enter and leave the team. \cite{wang2025n} considers extending AHT to settings where multiple learners are present. We assume fixed teams with a single learner.
\cite{paleja2021} considers AHT with humans, but focus on the benefits of incorporating explainable AI. 
Finally, self-play (SP) methods \cite{tesauroTDGammonSelfTeachingBackgammon1994} have been successful in zero-sum settings \cite{silver2016}, which are a type of AHT problem, but are ill-suited for cooperative AHT because they cannot coordinate effectively with non-SP agents \cite{carroll2019, hu_other-play_2020}.

\section{Preliminaries}
\label{sec:background}

\subsection{Multi-agent Reinforcement Learning}
\label{sec:marl}
We model our problem as a multi-agent Markov decision process (MMDP) defined by a tuple $\langle \mathcal{S}, \mathcal{N}, \{\mathcal{A}^i\}_{i \in \mathcal{N}}, p, r, \gamma \rangle$ \cite{boutilier1996}. Here,
$\mathcal{S}$ is the state space, $\mathcal{N}$ is the set of agents, $\mathcal{A}^i$ is the action space of agent $i$, and $\gamma \in [0, 1)$ is the discount factor. Let $\mathcal{A} \coloneqq \times_{i \in \mathcal{N}}\mathcal{A}^i$ be the joint action space. Then $p: \mathcal{S} \times \mathcal{A} \times \mathcal{S} \to [0,1]$ is the state transition function and $r: \mathcal{S} \times \mathcal{A} \times \mathcal{S} \to \mathbb{R}$ is the team reward function.
At time step $t$, each agent $i \in \mathcal{N}$ executes an action $a_t^i \in \mathcal{A}^i$ given the current state $s_t \in \mathcal{S}$, after which the system transitions to state $s_{t+1} \in \mathcal{S}$ and the team receives reward $r(s_t, \bm{a}_t, s_{t+1}$), where $\bm{a}_t \in \mathcal{A}$ is the joint action. Let $\pi^i: \mathcal{S} \times \mathcal{A}^i \to [0,1]$ be an individual policy for agent $i$ and $\pi(\bm{a}|s) \defeqn \prod_{i\in\mathcal{N}} \pi^i (a^i | s)$ be the resulting joint policy of all agents. The performance of a joint policy $\pi$ can be described by its action-value function,
\begin{equation}
\begin{split}
    Q^{\pi}(s, \bm{a}) \coloneqq & \mathbb{E}_{p,\pi} \Biggl[ \sum_{t=0}^\infty \gamma^{t} r(s_t, \bm{a}_t, s_{t+1}) \mid s_0 = s, \bm{a}_0 = \bm{a} \Biggr].
\end{split}
\label{eqn:q}
\end{equation}

\subsection{Generalized Policy Improvement}
\label{sec:gpu_sfs}
Consider a set of source tasks $\mathcal{R} = \{r_i\}_{i=1}^n$, where each task $r_i \in \mathcal{R}$ is defined as a linear reward,
\begin{equation}
    r_i(s,\bm{a},s') = \phi(s,\bm{a},s')^\intercal \mathbf{w}_i,
    \label{eqn:sf_r}
\end{equation}
where $\phi: \mathcal{S} \times \mathcal{A} \times \mathcal{S} \to \mathbb{R}^d$ is a function mapping to $d$ features and $\mathbf{w}_i \in \mathbb{R}^d$ is a weight vector specifying preferences over features. Following \cite{barreto_transfer_2018}, define the successor features (SFs) of a policy $\pi$ as,
\begin{equation}
    \psi^\pi(s, \bm{a}) \defeqn \mathbb{E}_{p,\pi} \left[ \sum_{t=0}^\infty \gamma^{t} \phi(s_t, \bm{a}_t, s_{t+1}) \mid s_0 = s, \bm{a}_0 = \bm{a} \right].
    \label{eqn:sf_psi}
\end{equation}
Then, the action-value function of $\pi$ on task $r_i$, $Q^\pi_{r_i}(s,\bm{a})$, is,
\begin{equation}
    Q^{\pi}_{r_i}(s, \bm{a}) = \psi^{\pi}(s,\bm{a})^\intercal \mathbf{w}_i.
    \label{eqn:sf_q}
\end{equation}

Now assume that we pretrain an agent on the set of source tasks $\mathcal{R}$ to generate a set of optimal policies $\Pi = \{ \pi_i^* \}_{i=1}^n$, where $\pi_i^*$ is an optimal policy for task $r_i \in \mathcal{R}$. Given a new target task $r_{n+1} = \phi(s,\bm{a},s')^\intercal \mathbf{w}_{n+1}$, we can use a GPI policy, $\pi'$, defined as,
\begin{equation}
    \pi'(s) \in \argmax_{\bm{a} \in \mathcal{A}} \max_{\pi \in \Pi} Q^\pi_{r_{n+1}} (s,\bm{a}),
    \label{eqn:gpi}
\end{equation}
to perform no worse than any policy in $\Pi$ on this task. If we compute the set of SFs associated with each policy in $\Pi$, $\Psi = \{ \psi^{\pi_i^*} \}_{i=1}^{n}$, we can compute the set $\{ Q_{r_{n+1}}^{\pi^*_i} \}_{i=1}^n$ using \Cref{eqn:sf_q} and implement $\pi'$ with no additional learning on the new target task. If additional learning is allowed, we can use $\pi'$ to quickly optimize a policy for $r_{n+1}$, for example using SFQL (Algorithm 3 in \cite{barreto_successor_2018}).
We use GPI as a framework for ZSC in AHT.

\subsection{Difference Rewards}
\label{sec:diff-rew}

In cooperative MARL, multiple agents interact within a shared environment to achieve a common goal, represented as a shared, team reward. A core challenge in this setting is the multi-agent credit assignment problem: determining which agent(s) were responsible for the reward resulting from their collective actions. Difference rewards offer a solution to this by approximating an individual agent's contribution to the team reward \cite{proper2012,castellini2022}. Instead of a team reward signal, each agent computes their individual difference reward and optimizes a policy to maximize its expected return. More formally, the difference reward for agent $i$, $\Delta r^i$, is defined as,
\begin{equation}
\label{eqn:dr}
    \Delta r^i \left( s, \bm{a}, s' \right) \defeqn r(s, \bm{a}, s') - \mathbb{E}_{b^i \sim \pi^{i}}\left[r\left(s, \langle \bm{a}^{-i}, b^i \rangle, s' \right) \right],
\end{equation}
where $\bm{a}^{-i}$ is the joint action of all agents other than $i$.
We extend the concept of difference rewards to the AHT setting so the learner can overcome the credit assignment problem.


\section{Problem Formulation}
\label{sec:formulation}

We modify the general formulation of MMDPs for AHT as follows. Let $\learner \in \mathcal{N}$ be the \emph{learner} (i.e., the agent whose policy we aim to optimize) and $\mathcal{N}_u = \mathcal{N} \setminus \{\learner\}$ be the complementary set of all teammates (i.e., uncontrolled agents). We assume that each teammate follows a fixed policy, which is unknown to the learner. Teammate policies may be suboptimal with respect to the team reward $r$ and the ad hoc team considered due to, e.g., the teammates being trained for a different task or with different teammates, or being humans and having inherent biases towards different goals. We formally define this model as an ad hoc MMDP.

\begin{definition}[Ad Hoc MMDP]
An ad hoc MMDP is defined by a tuple $M \defeqn \langle \mathcal{S}, \mathcal{N}, \learner, \{\mathcal{A}^i\}_{i \in \mathcal{N}}, p, r, \{\pi^i\}_{i \in \mathcal{N}_u}, \gamma \rangle$, where $\learner \in \mathcal{N}$ is the learner, $\mathcal{N}_u = \mathcal{N} \setminus \{\learner\}$ is the complementary set of teammates, and $\pi^i: \mathcal{S} \times \mathcal{A}^i \to [0,1]$ is the fixed policy of teammate $i$.
\end{definition}

We assume that the reward function, $r$, is non-negative. Note that any bounded reward can be transformed into a non-negative reward because scalar addition renders the reward to be policy invariant. We refer to an ad hoc MMDP $M$ as an ad hoc team. The performance of a learner policy $\pi^\learner$ in ad hoc team $M$ can be described by its action-value function,
\begin{equation}
\begin{split}
    Q^{\pi^\learner,\pi^{-\learner}}(s, a^\learner) \coloneqq \mathbb{E}_{p,\pi^\learner,\pi^{-\learner}} \Biggl[ \sum_{t=0}^\infty \gamma^{t} r(s_t, \bm{a}_t, s_{t+1}) \mid s_0 = s, a^\learner_0 = a^\learner \Biggr],
\end{split}
\label{eqn:q_i}
\end{equation}
where $\pi^{-\learner}$ is the joint policy of all teammates induced by $\{ \pi^i \}_{i\in\mathcal{N}_u}$. Our objective is to compute an optimal policy, $\pi^{\learner^*}$, which satisfies,
\begin{equation}
\label{eqn:pi_learner}
    Q^{\pi^{\learner^*}, \pi^{-\learner}} (s,a^\learner) \coloneqq \max_{\pi^\learner} Q^{\pi^{\learner}, \pi^{-\learner}}( s,a^\learner),
\end{equation}
for all $s \in \mathcal{S}$ and $a^\learner \in \mathcal{A}^i$. Optimizing a learner policy for an ad hoc team $M$ is equivalent to solving a single-agent Markov decision process with a transition function $\tilde{p}$ that captures the impact of teammate policies $\pi^{-\learner}$.

Assume we are given a partially specified ad hoc team $M_{\setminus \mathcal{N}_u} = \langle \mathcal{S}, \cdot, \learner, \mathcal{A}^\learner, p, r, \cdot, \gamma \rangle$ and $m$ possible ad hoc teammates $\{ \langle \mathcal{A}^i, \pi^i \rangle \}_{i=1}^m$. Then let $\mathcal{M}$ be the set of possible ad hoc teams induced by those teammates. We formalize our ZSC in AHT problem as follows.

\begin{problem}[Zero-shot Coordination for AHT]
\label{prob:zsc}
Let $\mathcal{M}_0 = \{ M_i \}_{i=1}^n \subseteq \mathcal{M}$ be a given set of source ad hoc teams with which the learner can pretrain. Our objective is to synthesize an optimal learner policy $\pi^{\learner^*}_{n+1}$ for a new ad hoc team $M_{n+1} \in \mathcal{M} \setminus \mathcal{M}_0$ by leveraging pretraining with $\mathcal{M}_0$ but with no online learning with the new team $M_{n+1}$.
\end{problem}


\section{Our Method}
\label{sec:method}

\subsection{Key Ideas: Generalized Policy Improvement and Difference Rewards}
\label{sec:ideas}

We address the AHT problem defined in \Cref{prob:zsc} through two key ideas. First, we use a GPI policy of the form in \Cref{eqn:gpi} to \emph{dynamically} leverage a library of pretrained learner policies to coordinate with a new ad hoc team with no online learning. By dynamic, we mean that GPI allows the learner to use different pretrained policies throughout the execution of a single episode, rather than being restricted to only using the best-matching pretrained policy (as in type-based methods) or a single robust policy (as in robust pretraining AHT methods). This ability to dynamically leverage policies is important, for example, in scenarios where a learner must use multiple pretrained skills to complete a task. Furthermore, this approach does not require online inference.

We are also motivated by the fact that a GPI policy guarantees improvement over each library policy in single-agent zero-shot transfer settings where dynamics are fixed and rewards are changed \cite{barreto_successor_2018}. However, in our setting, new ad hoc teammates induce new dynamics due to their (potentially) different policies, while our team rewards are fixed. These new dynamics make the set of pretrained SFs no longer valid, which prevents instant evaluation of the value functions for the new task that GPI requires. To ensure policy improvement, one would need to perform policy evaluation for \emph{each pretrained policy} with respect to the new ad hoc team, which requires many online samples.

\begin{figure}[t]
    \centering
    \includegraphics[width=0.75\linewidth]{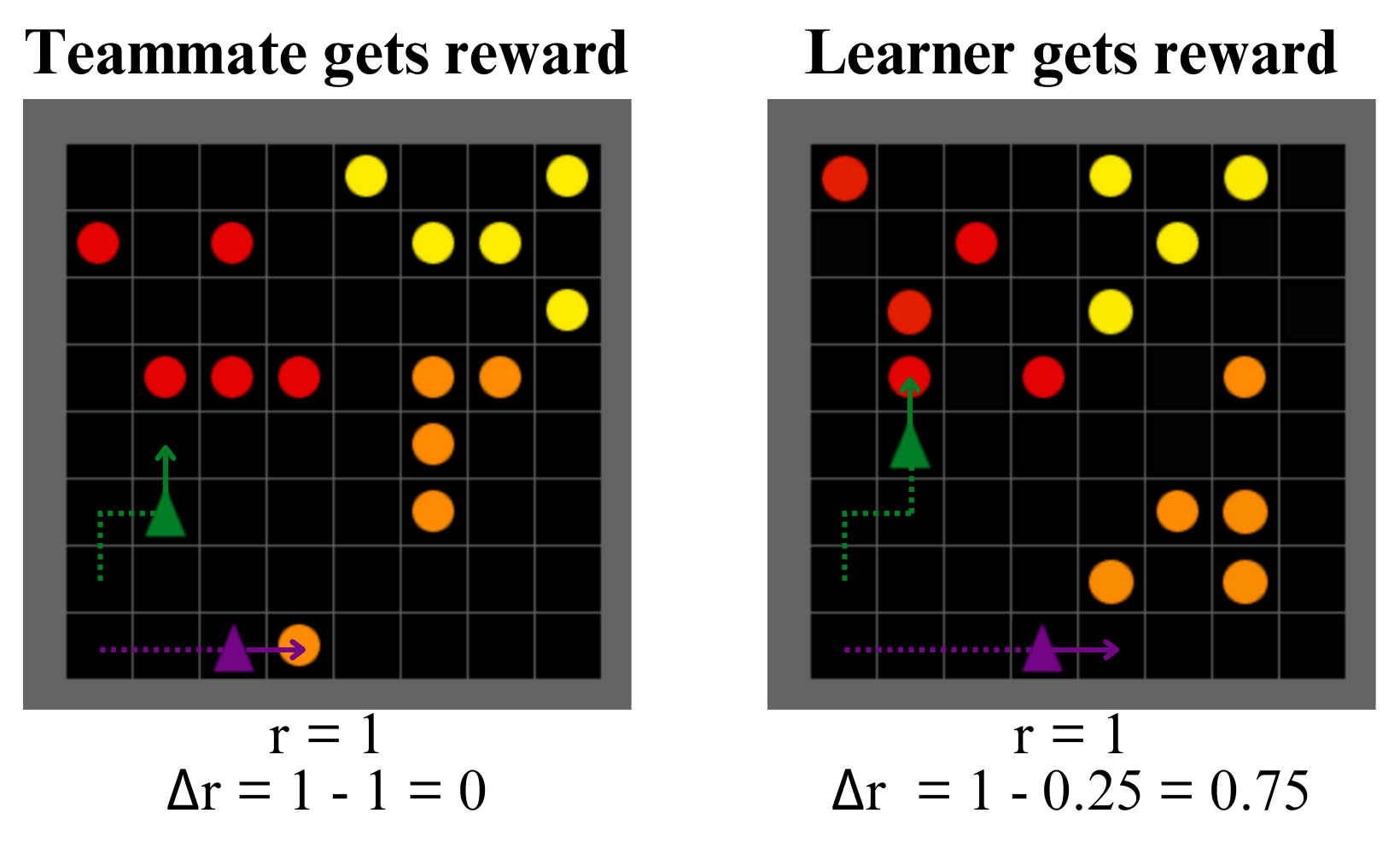}
    \caption{Difference rewards allow the learner to estimate its own contribution to the team reward. We illustrate this in a foraging example where collecting any object gives a team reward of 1. (left) the teammate action results in the reward and (right) the learner action results in the reward. While the team reward is the same in both scenarios, $\Delta r$ reflects the learner's contribution. We use this idea to reduce the effect of the distribution shift induced by new teammates.}
    \label{fig:dr-example}
\end{figure}

We address this issue through our second idea---instead of having GPI operate over value functions of pretrained policies evaluated with respect to the team reward, we have it operate over value functions with respect to the learner's difference rewards. Note that these value functions still assume teammate dynamics associated with the ad hoc teams used during pretraining; that is, we do not correct them to account for the new ad hoc team and therefore policy improvement is still not guaranteed. However, we hypothesize that evaluating with respect to the learner's difference rewards emphasizes contributions of the learner towards the team reward (while correspondingly de-emphasizing contributions of teammates), which will then reduce the impact of the distribution shift induced by the new ad hoc team on the actions selected by the GPI policy. This idea is illustrated in \Cref{fig:dr-example}. 

\subsection{An Algorithm for ZSC in AHT}
\begin{figure}[t]
    \centering
    \includegraphics[width=\linewidth]{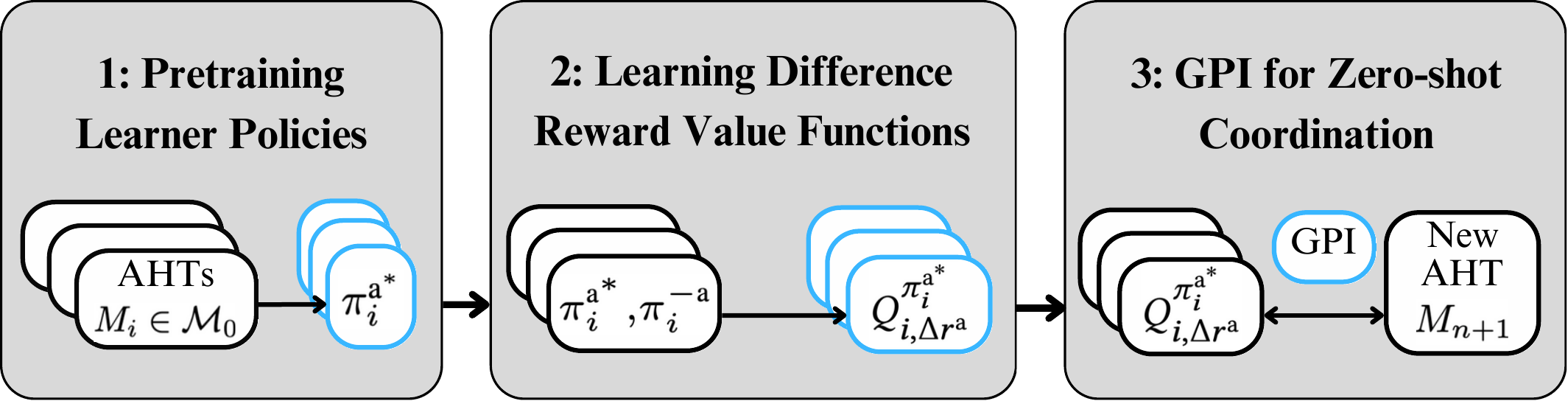}
    \caption{Our method, GPAT, addresses ZSC in AHT through 3 steps: (1) pretraining learner policies, (2) learning difference reward value functions to address the distribution shift induced by new teammates, and (3) dynamically leveraging this library at every time step by applying a GPI policy.}
    \label{fig:aht}
\end{figure}

Based on the ideas presented in \Cref{sec:ideas}, we propose an algorithm, \underline{GP}I for \underline{A}d Hoc \underline{T}eaming (GPAT), to address ZSC in AHT. Our algorithm is composed of three primary steps, visualized in \Cref{fig:aht} and discussed below. Pseudocode is provided in \Cref{alg:dr}. We consider two settings, one where the team reward $r$ can be modeled as a linear reward and the more general case of any team reward $r$. For the linear reward setting, we assume the features $\phi$ are fixed and heuristically defined such that $r$ can be modeled using \Cref{eqn:sf_r} through a weight vector $\mathbf{w}$. Our algorithm can be extended to incorporate feature and weight learning, as in \cite{barreto_fast_2020,hansen_fast_2020}.

\begin{algorithm}[t]
\caption{Generalized Policy Improvement for Ad Hoc Teaming}
\label{alg:dr}
\smaller
\begin{algorithmic}[1]
    \Statex \hspace{-\algorithmicindent}\hspace{-1.5mm} \textbf{Step 1:} Pretraining learner policies
    \Require $\Pi^{-\learner} \equiv \{\pi^{-\learner}_1, \dots, \pi^{-\learner}_n\}$
    \For{team $\in 1, \dots , i$}
        \If{linear reward}
        \State Optimize $\pi^\learner_i$ with $\pi^{-\learner}_i$ using SFQL or SFDQN
        \ElsIf{general reward}
        \State Optimize $\pi^\learner_i$ with $\pi^{-\learner}_i$ using any RL algorithm
        \EndIf
    \EndFor
\end{algorithmic}
\begin{algorithmic}[1]
    \Statex \hspace{-\algorithmicindent}\hspace{-1.5mm} \textbf{Step 2:} Learning difference reward value functions
    \Require $\Pi^{\learner} \equiv \{\pi^{\learner}_1, \dots, \pi^{\learner}_n\}, \Pi^{-\learner}, r(s, \bm{a}), \phi(s, \bm{a})$
    \For{team $\in 1, \dots , i$}
    \For{timestep $\in 1, \dots , T_{DR}$}
        \State $s \gets \texttt{env.reset()}$
        \While{not done}
            \State $\bm{a} \gets \pi^\learner_i, \pi^{-\learner}_i$
            \State $s', r \gets$ \texttt{env.step}($\bm{a}$)
            \State $\Delta r \gets \texttt{compute\_dr}(r, a^{-\learner})$ \Comment{\Cref{eqn:dr}}
            \If{linear reward}
            \State $\mathcal{D} \gets \Delta r, \phi(s, \bm{a})$
            \ElsIf{general reward}
            \State $\delta \gets \Delta r + \gamma Q_{i,\Delta r}(s', \pi^{\learner}(s')) - Q_{i,\Delta r}(s, a^{\learner})$
            \State $\theta \gets \theta + \alpha \delta \nabla_{\theta} Q_{i,\Delta r}(s, a^{\learner})$
            \EndIf
        \EndWhile
    \EndFor
    \If{linear reward}
        \State $w_{i,\Delta r} \gets \texttt{least\_squares}(\mathcal{D})$ \Comment{\Cref{eqn:sf_r}}
    \EndIf
    \EndFor
\end{algorithmic}
\begin{algorithmic}[1]
    \Statex \hspace{-\algorithmicindent}\hspace{-1.5mm} \textbf{Step 3:} GPI for zero-shot coordination
    \Require $\bm{Q}_{\Delta r^\learner} \equiv \{Q_{1, \Delta r^\learner}, \dots, Q_{n, \Delta r^\learner}\}$, $\pi^{-\text{a}}_{n+1}$
        \State $s \gets \texttt{env.reset()}$
        \While{not done}
            \State $a^\learner \gets \argmax_{b} \max_i Q^{\pi^{\learner^*}_i}_{i, \Delta r^\learner} (s, b) $ \Comment{\Cref{eqn:gpi-dr}}
            \State $\bm{a} \gets (a^\learner, \pi^{-\learner} (s))$
            \State $s, r \gets \texttt{env.step}(\bm{a})$
        \EndWhile
\end{algorithmic}
\end{algorithm}

\subsubsection*{Step 1: Pretraining Learner Policies.}
Given a set of source ad hoc teams $\mathcal{M}_0$, we first optimize a learner policy for each source ad hoc team. The output of this step is a library of learner policies $\Pi^\learner = \{ \pi^{\learner^*}_i \}_{i=1}^n$, where $\pi^{\learner^*}_i$ is the optimal learner policy for ad hoc team $M_i \in \mathcal{M}_0$. Any single-agent RL algorithm can be used for this step. In this work, we used Q-Learning with SFs (SFQL, Algorithm 3 in \cite{barreto_successor_2018}) for linear reward settings. This process produces a set of optimal learner SFs $\Psi^\learner = \{ \psi^{\pi^{\learner^*}_i, \pi^{-\learner}_i} \}_{i=1}^n$, where the learner SFs for learner policy $\pi^\learner$ in ad hoc team $M_i$ are defined as,
\begin{equation}
\begin{split}
    \psi^{\pi^\learner, \pi^{-\learner}_i} (s, a^\learner) = \mathbb{E}_{p,\pi^\learner,\pi^{-\learner}_i} \Biggl[ \sum_{t=0}^\infty \gamma^{t} \phi(s_t, \bm{a}_t, s_{t+1})
    \mid s_0 = s, a^\learner_0 = a^\learner \Biggr],
\end{split}
\label{eqn:sf_psi_learner}
\end{equation}
where $\pi^{-\learner}_i$ is the joint policy of all teammates in ad hoc team $M_i$. Following \Cref{eqn:sf_psi_learner}, we use these learner SFs to define corresponding action-value functions for policies in $\Pi^\learner$ as,
\begin{equation}
    Q^{\pi^\learner, \pi^{-\learner}_i} (s, a^\learner) = \psi^{\pi^\learner,\pi^{-\learner}_i} (s,a^\learner)^\intercal \mathbf{w}.
    \label{eqn:sf_q_learner}
\end{equation}
We refer to $\psi^{\pi^\learner, \pi^{-\learner}_i}$ as $\psi^{\pi^\learner}_i$ and $Q^{\pi^\learner, \pi^{-\learner}_i}$ as $Q^{\pi^\learner}_i$ hereafter to simplify notation, where $\psi^{\pi^\learner}_i$ and $Q^{\pi^\learner}_i$ are the SFs and action-value function for learner policy $\pi^\learner$ in ad hoc team $M_i \in \mathcal{M}$. We used PPO \cite{ppo} for general reward settings.

\subsubsection*{Step 2: Learning Difference Reward Value Functions.}
Given the library of pretrained learner policies $\Pi^\learner$, we now perform policy evaluation with respect to the learner's difference reward $\Delta r^\learner$, rather than the team reward $r$. Because the learner policies are deterministic, we assume a uniform learner policy when computing the difference rewards as in \cite{wolpert2001optimal}. The output of this step is a set $\mathcal{Q}_{\Delta r^\learner}^\learner = \{ Q_{i, \Delta r^\learner}^{\pi^{\learner^*}_i} (s, a^\learner) \}_{i=1}^n$, where $Q_{i, \Delta r^\learner}^{\pi^\learner}$ is the value function of policy $\pi^{\learner}$ with respect to the learner's difference reward in $M_i \in \mathcal{M}$.

\emph{Linear Reward Setting.}
We model the learner's difference reward as $\Delta r^\learner \left( s, \bm{a}, s' \right) = \phi(s,\bm{a},s')^\intercal \mathbf{w}_{\Delta r^\learner}$. Given this model, $\Psi^\learner$, and rollouts from the optimal learners in each ad hoc team $M_i \in \mathcal{M}_0$, we can now approximate $\mathcal{Q}_{\Delta r^\learner}^\learner$ by simply estimating $\mathbf{w}_{\Delta r^\learner}$ with linear regression and using \Cref{eqn:sf_q_learner}. We used Step 2 from \Cref{alg:dr} for this work, and show that a sufficiently accurate $\mathbf{w}_{\Delta r^\learner}$ can be learned in \textbf{as few as 10 episodes} in \Cref{sec:experiments}.
Note that this policy evaluation step can be performed during Step 1 using the same sampled experiences when using linear rewards---we simply separate this step for presentation purposes.

\emph{General Reward Setting.}
Any policy evaluation method can be used to estimate $\mathcal{Q}_{\Delta r^\learner}^\learner$ in the general reward setting. We use a TD-learning approach outlined in Step 2 from \Cref{alg:dr} for this work, which is a simplified version of fitted Q-iteration (FQI) algorithms \cite{Riedmiller2005NeuralFQ,NeumannFQI2008}. This process can be computationally expensive, but allows one to model a broader class of rewards.

\subsubsection*{Step 3: GPI for Zero-shot Coordination.}
We finally use a GPI policy for ZSC in a target (new) ad hoc team $M_{n+1}$. Given $\mathcal{Q}_{\Delta r^\learner}^\learner$, we define the GPI policy for the learner as,
\begin{equation}
\label{eqn:gpi-dr}
    \pi^{\learner} (s) \in \argmax_{a^{\learner} \in \mathcal{A}^\learner} \max_{i \in \{1, \dots, n \} } Q_{i, \Delta r^\learner}^{\pi^{\learner^*}_i} (s, a^\learner).
\end{equation}


\section{Experiments}
\label{sec:experiments}

\begin{table*}[t]
    \centering
    \smaller
    \caption{Foraging experiment descriptions. The reward weights and objects collected vectors correspond to [red, orange, yellow]. Skills from pretrained learners that are useful for the new AHT, and used by the oracle, are bolded. We design experiments to test 3 scenarios with differing percentages of useful pretrained skills.}
    \label{tab:foraging-exps}
    \begin{tabular}{p{1.7cm}ccccccc}
        \toprule
        \multicolumn{2}{c}{}
        & \multicolumn{ 2 }{ c }{ Source AHT 1 }
        & \multicolumn{ 2 }{ c }{ Source AHT 2 }
        & \multicolumn{ 2 }{ c }{ New AHT } \\ \cmidrule(lr){3-4} \cmidrule(lr){5-6} \cmidrule(lr){7-8}
        \multicolumn{2}{c}{}            
        & Teammate & \textbf{Learner}
        & Teammate & \textbf{Learner}
        & Teammate & Oracle
        \\ \midrule
        \multirow{3}{\linewidth}{\textit{Experiment 1 (50\% useful prior skills)}} & Reward weights & $[1.0, -0.5, -0.5]$ & $[1.0, 1.0, 1.0]$ & $[-0.5, 1.0, -0.5]$ & $[1.0, 1.0, 1.0]$ & $[-0.5, -0.5, 1.0]$ & $[1.0, 1.0, 1.0]$ \\
         & Objects collected & $[4.9, 0.0, 0.0]$ & $[0.1, 5.0, 4.9]$ & $[0.0, 4.6, 0.0]$ & $[5.0, 0.2, 4.9]$ & $[0.1, 0.1, 4.8]$ & $[4.9, 4.9, 0.1]$ \\
         & Preferred objects & red & \textbf{orange}, yellow & orange & \textbf{red}, yellow & yellow & \textbf{red, orange} \\ \hline
         \multirow{3}{\linewidth}{\textit{Experiment 2 (100\% useful prior skills)}} & Reward weights & $[0.0, 1.0, 1.0]$ & $[1.0, 1.0, 1.0]$ & $[1.0, 0.0, 1.0]$ & $[1.0, 1.0, 1.0]$ & $[-0.5, -0.5, 1.0]$ & $[1.0, 1.0, 1.0]$ \\
         & Objects collected & $[0.0, 4.9, 4.5]$ & $[5.0, 0.1, 0.5]$ & $[5.0, 0.0, 4.8]$ & $[0.0, 5.0, 0.2]$ & $[0.1, 0.1, 4.8]$ & $[4.9, 4.9, 0.1]$ \\
         & Preferred objects & orange, yellow & \textbf{red} & red, yellow & \textbf{orange} & yellow & \textbf{red, orange} \\ \hline
         \multirow{3}{\linewidth}{\textit{Experiment 3 (0\% useful prior skills)}} & Reward weights & $[0.0, 1.0, 1.0]$ & $[1.0, 1.0, 1.0]$ & $[1.0, 0.0, 1.0]$ & $[1.0, 1.0, 1.0]$ & $[1.0, 1.0, 0.0]$ & $[1.0, 1.0, 1.0]$ \\
         & Objects collected & $[0.0, 4.9, 4.5]$ & $[5.0, 0.1, 0.5]$ & $[5.0, 0.0, 4.8]$ & $[0.0, 5.0, 0.2]$ & $[4.8, 2.8, 0.2]$ & $[0.1, 2.2, 4.7]$ \\
         & Preferred objects & orange, yellow & red & red, yellow & orange & red, orange & \textbf{yellow} \\ \bottomrule
    \end{tabular}
\end{table*}

\subsection{Experimental Setup}
\label{sec:env}
\subsubsection*{Environments.} 
\begin{figure}[t]
    \centering
    \includegraphics[width=\linewidth]{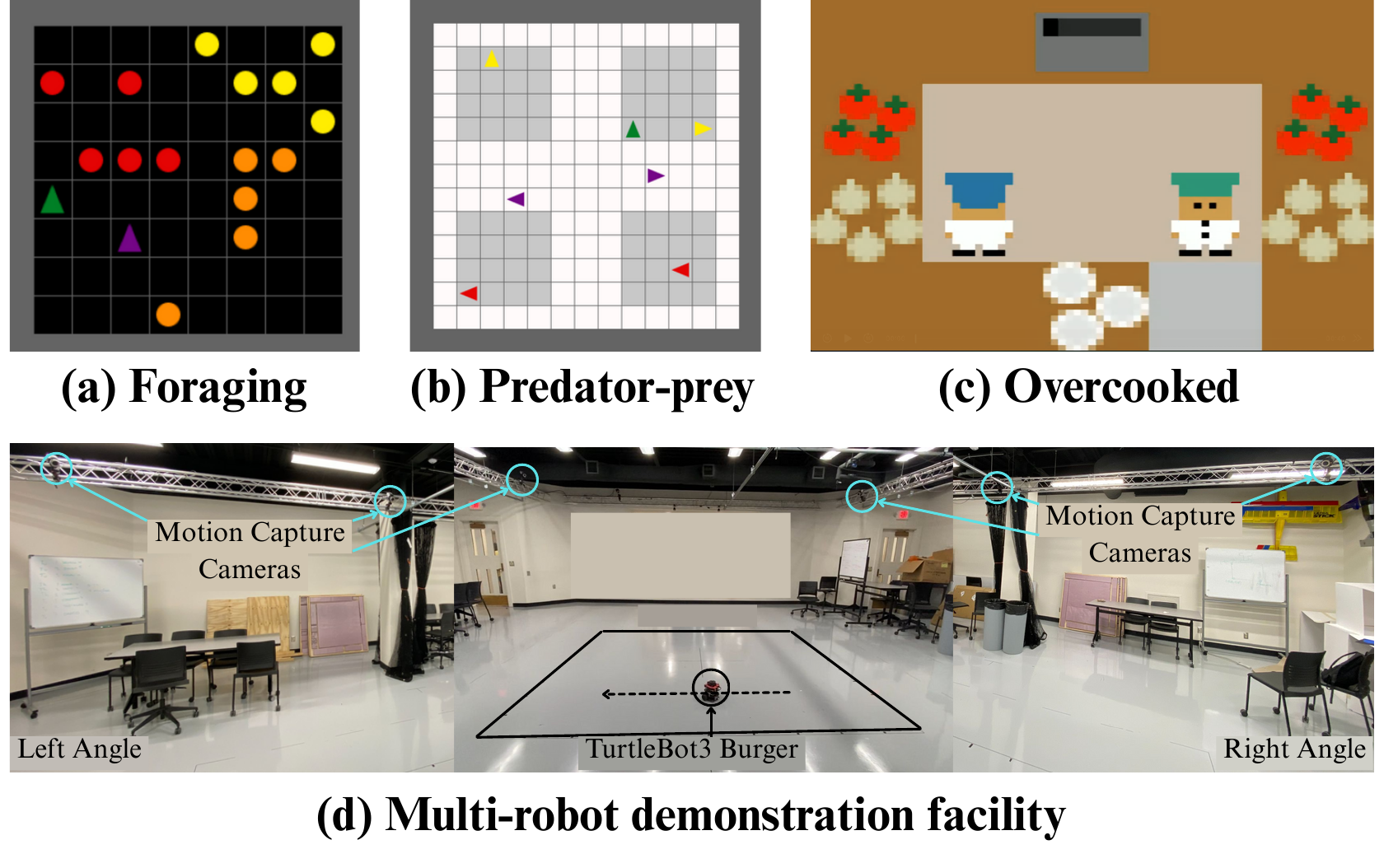}
    \caption{Multi-agent environments used in our experiments, where the learner agent is green. (a) Circles represent different types of objects agents aim to collect. (b) Agents aim to capture easy (yellow) and hard (red) prey. (c) Agents must coordinate to cook and serve a tomato onion soup. (d) Our multi-robot demonstration uses TurtleBot3 Burgers.}
    \label{fig:env}
\end{figure}

We empirically demonstrate GPAT's performance in a multi-agent foraging environment inspired by \cite{barreto_fast_2020, gu2022online}, a multi-agent predator-prey environment as in \cite{gu2022online, xing2023}, and Overcooked \cite{carroll2019, wang_too_2020}. The environments are illustrated in \Cref{fig:env}. We assume linear rewards for all environments, though we also consider a general reward setting for foraging within our ablation study.

The foraging environment has 2 agents in a team (one learner and one teammate) that aim to collect 3 types of objects (red, orange, and yellow) in an $8 \times 8$ grid. We define environment features $\phi$ as $1 \times 3$ vectors, where each element of $\phi$ is the number of objects of a given type that are collected in a state transition.
The team reward is then defined using $\textbf{w}=[1, 1, 1]$.
Following \cite{barreto_fast_2020}, our state representations are agent-centric and toroidal, such that the agent is always in the upper left corner and the grid is wrapped around the edges of the environment. The representation has 5 channels: one for each object type, one for the teammate, and one for walls. We cluster each object type in separate quadrants of the grid. Within each cluster, objects spawn in random locations at the start of each episode, and agents spawn in the lower left quadrant.

The predator-prey environment has 3 agents in a team (predators) that aim to capture 4 prey in a $13 \times 13$ grid. Prey randomly move within their shaded regions. We consider easy prey (yellow), which can be captured by a single predator, and hard prey (red), which must be captured by two predators. We define environment features and state representations similar to those in the foraging environment, with the team reward defined using $\textbf{w}=[1, 1, 1, 1]$.

The Overcooked environment has 2 agents in a team that aim to cook and deliver soups in a $2\times 3$ room---we use the Cramped Room layout, modified to include an additional soup ingredient, as in \cite{yu2023learning}, to increase coordination complexity (frequent interactions due to the small room size) and task complexity, as recommended in \cite{wang2024zsc}.
The agents must place an onion in the pot, place a tomato in the pot, start up the stove, wait for the soup to finish cooking, plate the soup, and finally deliver the soup. Following prior work \cite{yu2023learning}, we define environment features to consider 5 rewarding events (potting onion, potting tomato, picking up a dish, picking up the soup, delivering the soup) and define
the team reward as $\mathbf{w} = [3, 3, 3, 5, 20]$.
We use the default fully-observable state representation from \cite{carroll2019}.

\subsubsection*{Teammate Policies.}
For the foraging environment, we consider two source ad hoc teams in each experiment, where each experiment considers different combinations of source and new target ad hoc teams.
We design teammate policies to have different preferences for which object types (colors) they collect---the ideal learner collects the objects ignored by its teammate.
We optimize teammate policies using SFQL \cite{barreto_fast_2020}.
\Cref{tab:foraging-exps} defines our source and new target ad hoc teams, including statistics for the average number of object types each teammate collects in an episode.
We also include statistics for the resulting pretrained learners and an oracle learner that is able to train with the new target ad hoc team.

For the predator-prey environment, we consider two source ad hoc teams, where the teammates use heuristic greedy policies designed so that the agents have preferences for which prey they target---similar to the foraging environment, the ideal learner targets prey ignored by its teammates.

The teammate in Overcooked follows heuristic greedy policies designed such that the agent has preferences for performing specific tasks for the recipe.
Thus, the ideal learner performs the complementary tasks so the soup recipe can be completed and delivered.

\subsubsection*{Learner Policies.}
For the foraging and predator-prey environments, the learner policies were trained using SFQL \cite{barreto_fast_2020} with simple multilayer perceptron networks with two hidden layers of sizes 64 and 128. We used $\epsilon$-greedy exploration during training with $\epsilon=0.1$. Updates were done with batch sizes of $10$ with a discount factor of $\gamma=0.95$ and a learning rate of $\alpha = 3\times 10^{-5}$. All policies were trained for $2, 500, 000$ timesteps.
For the more complex Overcooked environment, we implemented an SF-DQN algorithm using the Stable Baselines3 DQN algorithm \cite{stable-baselines3}. We used the default hyperparameters, except for $\gamma=0.95$, and $2, 500, 000$ timesteps. Code for all experiments and environments is publicly available\footnote{\url{https://github.com/apurl1/gpat}}.

\subsubsection*{Baselines.}

We implement GPAT with linear rewards in our experiments (unless otherwise noted) and learn $w_{\Delta r}$ with 10 episodes, based on preliminary experiments ranging from 10-100 evaluation episodes.
We compare GPAT to the following baselines:
\begin{itemize}
    \item \emph{Oracle:} We train a learner from scratch (using SFQL or SFDQN) with the new target AHT to represent an oracle learner.
    \item \emph{Robust:} We pretrain a learner with all source AHTs to represent robust ZSC methods like \cite{hu_other-play_2020, zhao_maximum_2022, yu2023learning}, where we randomly sample the training team as in \cite{yu2023learning}. We train robust learners (using SFQL or SFDQN) for the total timesteps used to train the entire learner policy libraries used by other methods. We assume the source AHTs are given to the learner, so we do not consider the source AHT generation process.
    \item \emph{PLASTIC:} We use the best pretrained learner policy from the library $\Pi^\learner$ to represent a best-case scenario for type-based methods like \cite{BARRETT-plastic, zhao_coordination_2022, zandOntheFly2022}. We skip the online inference phase and simply use the best pretrained policy to mimic a setting where inference is already performed and correct.
    \item \emph{MESH:} We implement a zero-shot version of \cite{zhao_coordination_2022}, which combines pretrained policies from $\Pi^\learner$ using a mixture of experts model. We use a uniform mixture since we do not allow online learning.
\end{itemize}

\begin{table*}[t]
    \centering
    \caption{Experiment results, showing IQM returns for GPAT compared to baselines \cite{BARRETT-plastic, yu2023learning, hu_other-play_2020}, measured using 1000 episodes with 10 replicates and 95\% confidence intervals. The percentage optimality with respect to the oracle is also reported in parentheses.}
    \label{tab:foraging-results}
    \begin{tabular}{cccccc}
        \toprule
        \multicolumn{1}{c}{}
        & \multicolumn{ 3 }{ c }{ Foraging }
        & \multicolumn{ 1 }{ c }{ Predator-Prey }
        & \multicolumn{ 1 }{ c }{ Overcooked } \\ \cmidrule(lr){2-4}
         & Experiment 1 & Experiment 2 & Experiment 3 & & \\ \midrule
         Oracle & $8.087 \pm 0.011$ & $8.082 \pm 0.011$ & $8.091 \pm 0.012$ & $2.547 \pm 0.008$ & $607.76 \pm 0.735$ \\
         \textbf{GPAT (ours)} & $\mathbf{7.755 \pm 0.014 (95.9\%)}$ & $\mathbf{7.635 \pm 0.019 (94.5\%)}$ & $5.641 \pm 0.024 (69.7\%)$ & $\mathbf{2.202 \pm 0.011 (86.5\%)}$ & $\mathbf{218.07 \pm 5.401 (35.9\%)}$ \\
         Robust & $5.998 \pm 0.018 (74.2\%)$ & $6.590 \pm 0.016 (81.5\%)$ & $\mathbf{6.573 \pm 0.016 (81.2\%)}$ & $2.077 \pm 0.011 (81.5\%)$ & $33.05 \pm 2.078 (5.4\%)$ \\
         PLASTIC & $6.438 \pm 0.010 (79.6\%)$ & $6.235 \pm 0.012 (77.1\%)$ & $6.238 \pm 0.013 (77.1\%)$ & $2.096 \pm 0.008 (82.3\%)$ & $9.69 \pm 0.328 (1.6\%)$ \\ 
         MESH & $4.654 \pm 0.032 (57.5\%)$ & $5.485 \pm 0.030 (67.9\%)$ & $5.930 \pm 0.024 (73.3\%)$ & $1.569 \pm 0.012 (61.6\%)$ & $0.0 \pm 0.0 (0.0\%)$ \\ \bottomrule
    \end{tabular}
\end{table*}

\begin{table*}[t]
    \centering
    \caption{Ablation results in the foraging environment, showing IQMs returns for GPAT with a linear reward (ours), general reward (GR), and without difference rewards (w/o DR), measured using 1000 episodes with 10 replicates and 95\% confidence intervals. We also report the percentage optimality and the percentage each pretrained policy is used.}
    \label{tab:ablation-results}
    \begin{tabular}{cccccccccc}
         \toprule
         \multicolumn{1}{c}{}
         & \multicolumn{3}{c}{Experiment 1}
         & \multicolumn{3}{c}{Experiment 2}
         & \multicolumn{3}{c}{Experiment 3} \\ \cmidrule(lr){2-4} \cmidrule(lr){5-7} \cmidrule(lr){8-10}
         & Return & $\% \pi_1$ & $\% \pi_2$ & Return & $\% \pi_1$ & $\% \pi_2$ & Return & $\% \pi_1$ & $\% \pi_2$ \\ \midrule
         GPAT (ours) & $\bm{7.755 \pm 0.014 (95.9\%)}$ & $56.5\%$ & $43.5\%$ 
                & $\bm{7.635 \pm 0.019 (94.5\%)}$ & $50.1\%$ & $49.9\%$
                & $5.641 \pm 0.024 (69.7\%)$ & $10.0\%$ & $90.0\%$ \\
         GPAT with GR & $7.266 \pm 0.022 (89.8\%)$ & $56.2\%$ & $43.8\%$ 
                 & $7.492 \pm 0.021 (92.7\%)$ & $45.2\%$ & $54.8\%$
                 & $\bm{6.282 \pm 0.024 (77.6\%)}$ & $5.6\%$ & $94.4\%$ \\
         GPAT w/o DR & $7.066 \pm 0.019 (87.4\%)$ & $34.7\%$ & $65.3\%$ 
               & $6.625 \pm 0.019 (82.0\%)$ & $72.4\%$ & $27.6\%$
               & $6.206 \pm 0.015 (76.7\%)$ & $22.6\%$ & $77.4\%$ \\ \bottomrule
    \end{tabular}
\end{table*}


\subsection{Results}
\label{sec:zsc-collect}


Our results show inter-quartile mean (IQM) evaluation returns for the learner with the new target partner with 95\% confidence intervals (CIs), following the recommendations from \cite{agarwal2021deep}. Higher IQM scores are better. The CIs are estimated using percentile bootstrap with stratified sampling with 1,000 bootstrap resamples. We saw similar trends when analyzing mean and median results (not shown), and thus only show IQM results because it is more robust to outliers than the mean and has less bias than the median \cite{agarwal2021deep}.

\subsubsection*{Foraging.} We evaluate GPAT in the three experiments shown in \Cref{tab:foraging-exps}. Experiment 1 requires the learner to leverage one of the two skills learned from each pretrained policy in order to optimally coordinate with the new teammate. Experiment 2 requires the learner to leverage all skills learned during pretraining. Experiment 3 contains no relevant skills in the learner's pretrained policies. \Cref{tab:foraging-results} shows results for these experiments. We see that GPAT (with a linear reward) outperforms all baselines (other than the oracle) for Experiments 1 and 2, likely because it is able to effectively use both pretrained skills.
In comparison, the Robust baseline is likely struggling to generalize its pretrained skills to the new out-of-distribution teammate. The PLASTIC baseline is limited to using the best-matching pretrained policy and thus can only implement one of the needed skills.
However, GPAT performs worse than both baselines in Experiment 3, likely due to the learner's pretrained library not containing the skill needed to coordinate with the new teammate (i.e., collecting yellow objects).
In principle, GPAT could perform similarly to PLASTIC, since it's library contains the best-matching policy used by PLASTIC---however, we see that GPAT is outperformed by PLASTIC, indicating that it occasionally picks the worse policy, likely due to inaccuracies incurred by the distribution shift induced by the new teammate. Note that our ablation results (shown in \Cref{tab:ablation-results}) show that GPAT with general rewards performs similarly to our baselines in Experiment 3. We discuss this point more in the ablation discussion.
Overall, these experiments suggest that GPAT can effectively achieve ZSC when its library has at least some relevant skills, but can struggle when there are no relevant skills in the library.

\subsubsection*{Predator-Prey.} In the predator-prey experiment, the first pretrained learner policy learns to target the top-right and bottom-right prey, while the second pretrained learner policy learns to target the top-left and bottom-left prey. To coordinate optimally with the new AHT, the learner must target the top-left and bottom-right prey, thus combining both pretrained skills. As shown in \Cref{tab:foraging-results}, we see that GPAT outperforms our baselines due to its ability to correctly use both pretrained policies as needed. Thus, GPAT can coordinate well with multiple teammates and handle dynamic environments.

\begin{figure}[t]
    \centering
    \includegraphics[width=0.95\linewidth]{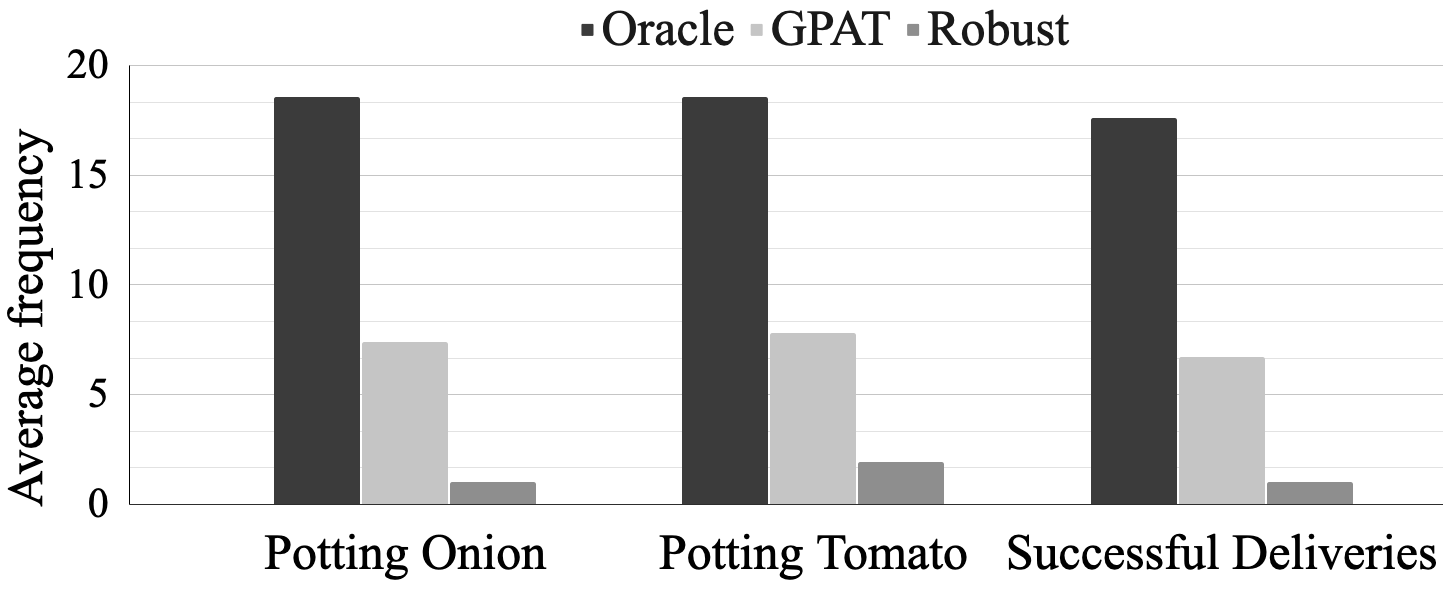}
    \caption{The average number of times the learner performs various subtasks over an episode, measured using 1000 episodes with 10 replicates. The Robust learner struggles to pot onions and tomatoes, despite learning how to optimally perform both during pretraining.}
    \label{fig:rob-overcooked}
\end{figure}

\begin{figure}[t!]
    \centering
    \includegraphics[width=0.7\linewidth]{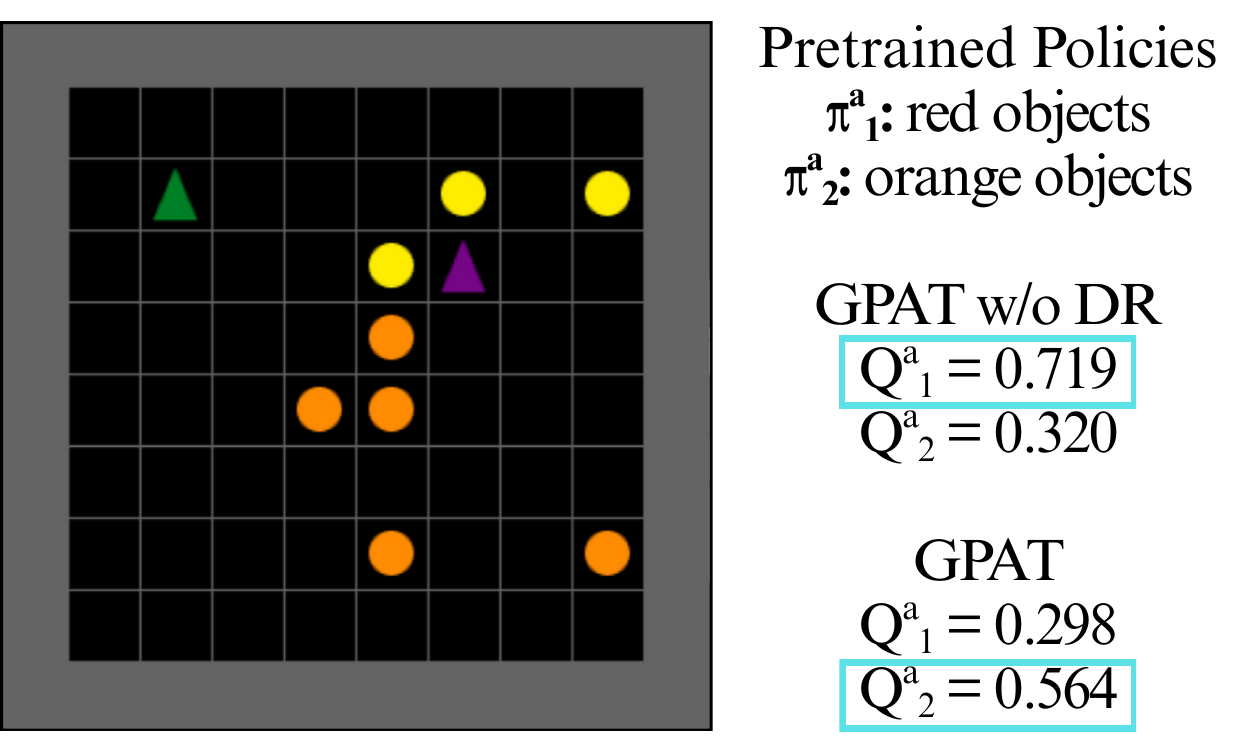}
    \caption{Q-values for a state from Experiment 2 in the foraging environment, normalized by the maximum for each method. Without difference rewards, the learner is unable to switch policies once all red objects have been collected.}
    \label{fig:switch}
\end{figure}

\begin{figure*}[t]
    \centering
    \includegraphics[width=0.92\linewidth]{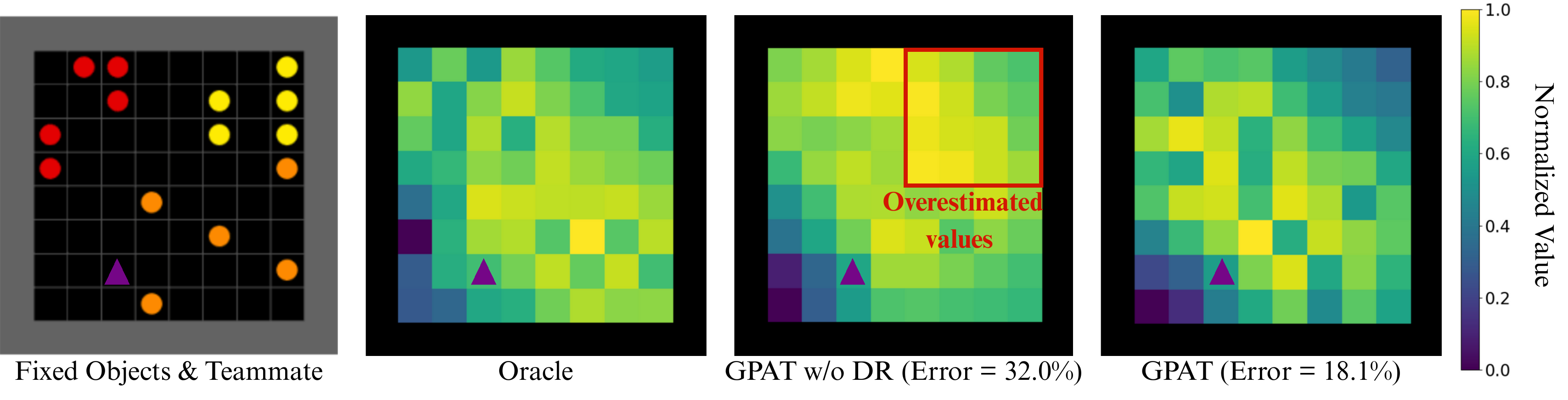}
    \caption{Normalized value maps for a state (left) in the foraging environment for Experiment 2, where we show the value of the learner being in each grid position and the average percent error of the values relative to the oracle's in parenthesis. Without difference rewards, the learner overestimates the value of being near yellow objects (the new teammate's preferred object) and has a higher error than GPAT.}
    \label{fig:vmaps}
\end{figure*}

\subsubsection*{Overcooked.} In Overcooked, the first pretrained learner trains with an onion-preferring teammate to learn to pot tomatoes, cook the soup, plate the soup, and deliver the soup. The second pretrained learner trains with a tomato-preferring teammate to learn to pot onions, cook the soup, plate the soup, and deliver the soup. To coordinate optimally with the new dish-preferring teammate, the learner must pot both onions and tomatoes before cooking the soup.
\Cref{tab:foraging-results} shows that GPAT significantly outperforms our baselines.
Compared to other environments, there is also a greater optimality gap between all methods and the oracle.
We hypothesize that these trends are due to the sequential nature of the task, where it is necessary to complete each step in the correct order, which thus requires more intelligent coordination than other environments.

We further investigate why the Robust learner struggles to perform well, despite converging to optimal returns during pretraining with library teammates. \Cref{fig:rob-overcooked} visualizes the number of times the learner pots onions or pots tomatoes, along with how many deliveries the team successfully made.
We observe that the Robust learner does not perform the required subtasks often enough, indicating that it may have overfit to the training teammates and does not know how to optimally use those skills with the new teammate.

\subsubsection*{Ablation Results.} \Cref{tab:ablation-results} shows results from an ablation study for GPAT. We used PPO from Stable Baselines3 \cite{stable-baselines3} to train learner policies with general rewards, and performed policy evaluation using TD-updates (see \Cref{alg:dr}) with 2500 episodes, based on preliminary experiments ranging from 100-5000 episodes. We see that GPAT with linear rewards outperforms general rewards in Experiments 1 and 2, likely due to easier reward learning. This, along with the much greater sample efficiency of learning $w_{\Delta r}$ compared to learning $Q_{\Delta r}$, is why we use GPAT with linear rewards in the main results. However, GPAT with general rewards outperforms linear rewards in Experiment 3, where it performs similarly to the Robust and PLASTIC baselines. Recall that in Experiment 3, the pretraining library does not contain any relevant skills for the new teammate. Thus, we hypothesize that the noisier policy learned by general rewards is actually advantageous in this scenario.
\balance
\Cref{tab:ablation-results} also shows that removing difference rewards results in one pretrained policy being used much more than the other---ideally, we would expect the learner to roughly select each pretrained policy an equal number of times in Experiments 1 and 2.
\Cref{fig:switch} shows an example state where GPAT w/o DR is unable to switch between policies.
Here, all red objects have been collected. Since the new teammate prefers yellow objects, the learner should switch to collecting orange (using $Q^\learner_2$).
However, GPAT without DR is unable to appropriately switch policies because it overestimates the value of $\pi^\learner_1$, since its value functions are with respect to the team reward and thus assume the teammate will collect the remaining objects.
Furthermore, \Cref{fig:vmaps} shows that removing difference rewards decreases performance, as it results in overestimated values near yellow objects (which is the new teammate's preference) due to the value function being with respect to the team reward.
We also see that, on average, GPAT w/o DR has a higher value error than GPAT. Thus, difference rewards results in better aligned values despite the new dynamics induced by the new teammate.

\subsection{Real World Multi-Robot Demonstration}
\label{sec:robot}
\begin{figure}[t!]
    \centering
    \includegraphics[width=0.8\linewidth]{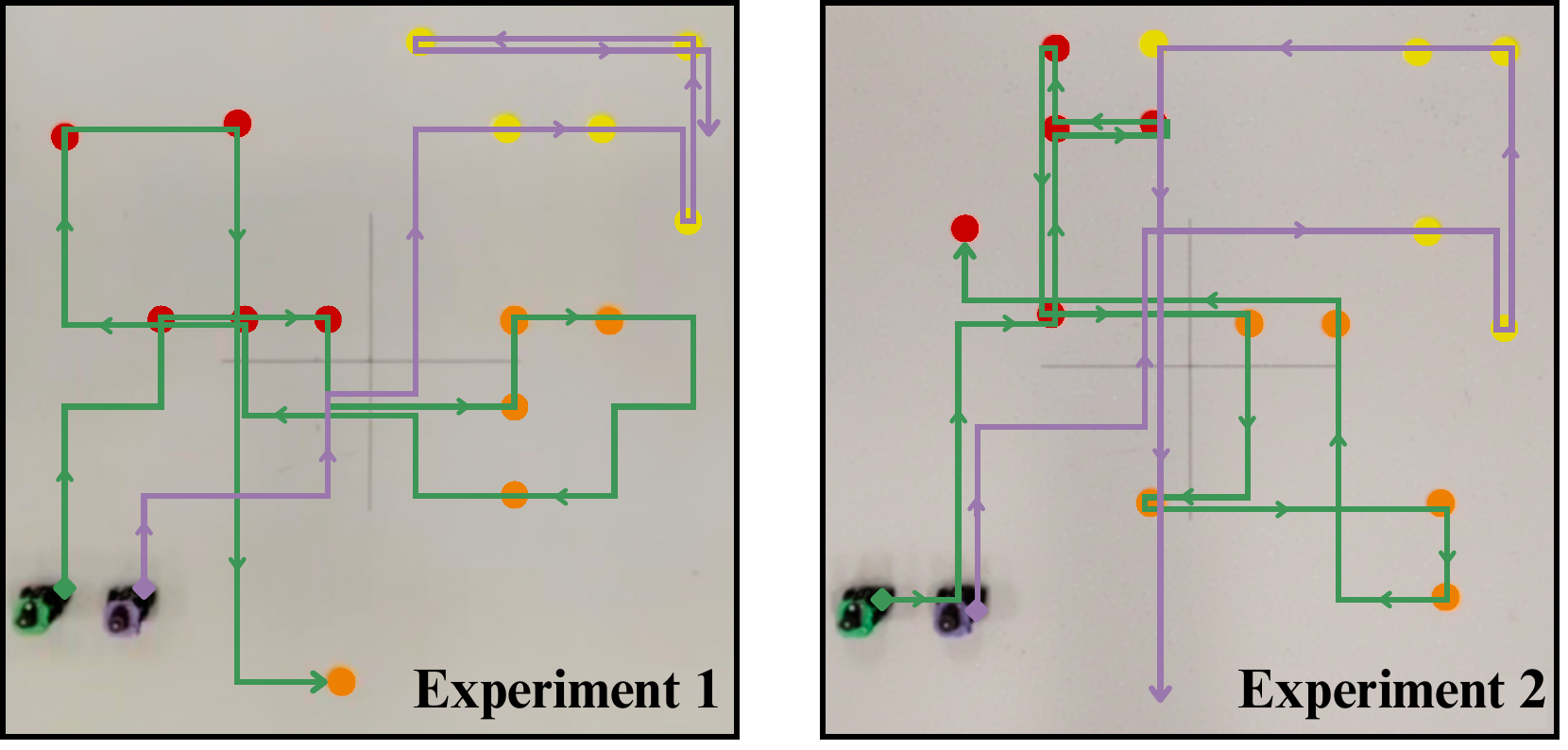}
    \caption{Multi-robot demonstration of the foraging environment. We overlay the trajectories taken by the robots for two experiments from \Cref{tab:foraging-exps} (green=learner, purple=teammate).}
    \label{fig:robot-demos}
\end{figure}

We also demonstrate our method in a real-world multi-robot setting using Robotis Turtlebot3 Burgers in the foraging environment. We employ two TurtleBot3 Burger robots, each running a separate instance of the Robot Operating System (ROS) and maintaining its own identification parameters. A Qualysis motion capture system, with four passive markers attached to the robots, provides precise localization within a 12 ft.$\times$~12 ft. grid-world. At every time step, we use GPAT to generate control commands, which are then transmitted to the robots through ROS. The trajectories the robots take are visualized in \Cref{fig:robot-demos}. We observe the learner robot (green) collects the red and orange objects while the teammate robot (purple) collects the yellow objects, as expected. A video illustrating these real-world experiments is available online.\footnote{Robot video: \url{https://youtu.be/sm6Y-AakSgA}}

\section{Conclusion}
\label{sec:conclusion}
In this work, we propose GPAT, a novel approach that leverages GPI and difference rewards for ZSC in AHT. In contrast to previous methods that use a single policy from a pretrained library with online inference or train a single policy robust to diverse teammates, we dynamically leverage the entire library of pretrained policies at every time step by applying a GPI policy. We address the impact of the distribution shift induced by new teammates on the GPI policy through difference rewards. We empirically demonstrate that this more exhaustive use of prior knowledge improves performance in three simulated environments relative to baselines. We also include a multi-robot demonstration of our method. Future directions include theoretical analysis, considering online adaptation, and considering partial observability.



\begin{acks}
This work was supported in part by ONR N00014-20-1-2249, a NASA grant awarded to the Illinois/NASA Space Grant Consortium, and a GAANN grant from the U.S. Department of Education.
\end{acks}



\bibliographystyle{ACM-Reference-Format} 
\bibliography{references, references-huy}

\end{document}